\documentclass[12pt]{article}
\pdfoutput=1
\usepackage{graphics}
\usepackage[dvips]{graphicx}
\usepackage{mathrsfs}
\usepackage{amssymb}
\usepackage{amsmath}
\usepackage{verbatim}
\usepackage{float}
\usepackage{slashed}
\usepackage{bbm}
\newcommand{\be}{\begin{equation}}
\newcommand{\ee}{\end{equation}}
\newcommand{\bea}{\begin{eqnarray}}
\newcommand{\eea}{\end{eqnarray}}

\def\4vol{{\int d^4x \sqrt{-g}}}

\def\beq{\begin{equation}}
\def\eeq{\end{equation}}
\def\bea{\begin{eqnarray}}
\def\eea{\end{eqnarray}}
\def\bitem{\begin{itemize}}
\def\eitem{\end{itemize}}
\def\ba{\begin{array}}
\def\ea{\end{array}}
\def\bal{\begin{align}}
\def\eal{\end{align}}
\def\bi{\begin{itemize}}
\def\ei{\end{itemize}}

\newcommand{\nc}{\newcommand}
\nc{\nt}{\tilde{N}}
\nc{\ra}{\rightarrow}
\nc{\lsim}{\begin{array}{c}\,\sim\vspace{-21pt}\\< \end{array}}
\nc{\gsim}{\begin{array}{c}\sim\vspace{-21pt}\\> \end{array}}

\nc{\ie}{{\it i.e.~}}          \nc{\etal}{{\it et al.~}}
\nc{\eg}{{\it e.g.~}}          \nc{\etc}{{\it etc.~}}
\nc{\cf}{{\it c.f.~}}

\nc{\LL}{L}
\nc{\vv}{\tilde{v}}

\title{
\vspace*{5mm} \Large\textbf{Discovering Nonstandard Higgs bosons in the $H\to ZA$ Channel Decay to Multileptons}
\vspace*{1.0cm}
\author{\textbf{Spencer Chang and Arjun Menon}\\
  \normalsize\emph{Institute of Theoretical Science, University of Oregon, Eugene, OR  97403, USA}\\
}
}                
\date{\today}
\begin{document}
\setcounter{page}{0}
\maketitle
\begin{abstract}
In this article we consider the possibility of observing nonstandard Higgs 
bosons in the $H \to Z A \to Z \tau^+ \tau^-$ channel. We present three 
benchmark scenarios in the
NMSSM where $H \to ZA$ is the dominant decay mode for one of the nonstandard 
Higgs bosons while the lightest CP-even Higgs is Standard Model like. Using the latest CMS multilepton analysis based on  
7~TeV LHC data, we put limits on the signal
cross-section, which constrain leptophilic scenarios.  
Projecting to future LHC analyses with improvements in background modeling,  we show that with $O(30)$ fb$^{-1}$ of data, such a multilepton analysis is very close to constraining our NMSSM benchmarks.  
As we illustrate with a toy model, for light $A$ masses, the large boost of the $A$ makes it inefficient to select two hadronic taus, since isolation and the transverse momenta are in tension.   This efficiency could be improved by including boosted di-tau jets as an object in future multilepton analyses.    
We also discuss  different methods to confirm this scenario by reconstruction of  the $m_H$ and $m_A$ masses.
In particular we consider the transverse mass distribution, collinear mass
distribution and an analytical solution using trial masses.
\end{abstract}

\thispagestyle{empty}
\newpage
\setcounter{page}{1}

\section{Introduction}
\label{sec:Intro}

Recently, ATLAS and CMS announced the discovery of  a particle consistent with the Standard Model Higgs \cite{ATLASHiggs, CMSHiggs}.  If the properties of this particle are confirmed to be that of a Higgs boson, this will be a major revolution in particle physics, as it would be the first fundamental scalar field proven to exist.  This discovery opens up the possibility for other fundamental scalars and thus motivates searches at the LHC for additional spin zero particles.  Moreover, in many theories beyond the Standard Model, {\it e.g.}~supersymmetry or two Higgs doublet models, such additional Higgs bosons are required and thus their discovery would be an important first step towards uncovering this new physics.

The phenomenology of such scalars has so far been incompletely explored.  In the minimal supersymmetric Standard Model, there is a well explored phenomenology of the heavy Higgs states $H, A$ with ongoing searches in $\tau^+ \tau^-$ \cite{ATLASHeavy, CMSHeavy} and potentially observable $b\bar{b}$ decays \cite{Carena:2012rw}.  However, scalars can also have extremely challenging signals, many of which were explored in the nonstandard Higgs scenario \cite{nonstandardreview}.  Thus, it is fruitful to continue to explore promising signals for new bosons.  

In this note, we point out an interesting signal with the potential to discover two new scalars.  The signal process is $H \to Z, A \to \ell^+ \ell^-, \tau^+ \tau^-$ which has been studied in the context of explaining LEP anomalies~\cite{Dermisek:2008id,Dermisek:2008uu}.  Such a signal is motivated by the recent CMS multilepton search \cite{CMS2} in particular its channels with an onshell leptonic $Z$.  Recently, the multilepton search has also been shown to be sensitive to Standard Model Higgs modes \cite{ContrerasCampana:2011aa},  flavor violating top decays into a Higgs $(t\to c h)$ \cite{Craig:2012vj}, and two Higgs doublet models \cite{RUTGERSfolk}. These previous analyses did not consider reconstructing hadronic taus and focused on an inclusive search using all the CMS multilepton channels.  Considering our signal's exclusive channels into hadronic taus provides an important additional probe which is sensitive to a substantial branching fraction of the signal and also allows a variety of mass reconstruction techniques given the limited number of neutrinos.  Admittedly, the substantial excess in the original CMS multilepton search \cite{CMS1} in these channels went down in \cite{CMS2}, but this signal still remains promising in future updates.   

In this paper, we analyze the prospects a CMS-like multilepton analysis has in discovering such a signal.  The outline of the rest of the paper is as follows.  In Sec.~\ref{sec:model}, we explore simple benchmarks  to realize such a signal in the Next-to-Minimal Supersymmetric 
Standard Model.  This provides an important existence proof in a motivated theory.   More optimistic scenarios exist as well in leptophilic two Higgs doublet models.  Those interested in the phenomenology can skip to Sec.~\ref{sec:limits}, where we estimate the efficiency the CMS multilepton search has on such a signal and use them to derive model-independent bounds on the signal rate.  In Sec.~\ref{sec:recon_comp}, we compare the utility of a variety of mass reconstruction techniques for determining the $H, A$ masses.  In Sec.~\ref{sec:concl}, we conclude and look at future directions.   In Appendix \ref{sec:mhmatest}, we give details for solving the neutrino momenta given trial masses for the two bosons. 

\section{A simple model for $H \to ZA$ signals}
\label{sec:model}

In this section we discuss the possibility of enhancing the coupling of 
a CP-even Higgs boson ($H$) to the Z gauge boson and a pseudo-scalar ($A$). We 
would like to develop a model where the $H$ decays predominantly to $ZA$
and the $A$ in turn decays into $\tau^+\tau^-$. This model has the possibility
of explaining the slight excesses observed by the CMS 
collaboration~\cite{CMS2} in 
the 4$\ell+0\tau_h$, 3$\ell+1\tau_h$ and $2\ell+2\tau_h$ channels with a reconstructed leptonic $Z$.   In these channels, the $\tau_h$'s indicate reconstructed one-prong 
hadronic decays. Therefore, to realize this scenario in a model we need the $A$ to
be reasonably light and the branching ratios of $H \to ZA$ and $A \to \tau^+\tau^-$ to be
significant.  In particular, we are interested in benchmarks where $m_{H} \sim 200$ GeV and $m_{A} \sim 10-100$ GeV, ensuring that the $Z$ decay is open and the $H$ can be produced with reasonable rates.  Such a decay is particularly difficult to realize in the Minimal Supersymmetric Standard Model due to the decoupling limit constraining both the mass of $A$ and the decay of $H\to Z A$ and thus, we have to turn to other theories.  

To demonstrate a realization of such a model we consider the Next-to-Minimal Supersymmetric 
Standard Model (NMSSM)~\cite{NMSSM1}.  The superpotential has the form
\bea
W = W_{\rm Yuk} + \lambda \hat H_u \hat H_d \hat S + \frac{\kappa}{3} \hat S^3
\label{eq:superpotential}
\eea
where $W_{\rm Yuk}$ are the usual Yukawa interactions and the hatted fields 
denote the chiral superfields. The corresponding soft supersymmetry breaking
terms are 
\bea
V_{\rm soft} = m_{H_u}^2 |H_u|^2 + m_{H_d}^2 |H_d|^2 + m_S^2 |S|^2 + \sqrt{2}
\left(m_\lambda S H_u H_d - \frac{m_\kappa}{3} S^3\right).
\label{eq:vsoft}
\eea
Here, we will follow the discussion and notation of Ref.~\cite{Dobrescu:2000yn}.  The relationships to the  
standard NMSSM notation of Ref.~\cite{Ellwanger:2004xm} are
$m_\kappa \equiv -\kappa A_\kappa/\sqrt{2}$ and $m_\lambda \equiv \lambda A_\lambda
/\sqrt{2}$. We will
work in the CP-even Higgs basis $(h_v^0, H_v^0,h_s^0)$ and CP-odd basis $(A_v^0,
A_s^0)$, where  
\bea
H_d^0 &=& \frac{1}{\sqrt{2}} [(v+h_v^0-iG^0) c_\beta - (H_v - i A_v^0) s_\beta] \\
H_u^0 &=& \frac{1}{\sqrt{2}} [(v+h_v^0+iG^0) s_\beta - (H_v + i A_v^0) c_\beta] \\
S &=& \frac{1}{\sqrt{2}} (s+h_s^0 + i A_s^0),
\eea
$v \sim 246$ GeV is the Higgs vacuum expectation value (VEV), $\tan \beta = (v_u/v_d)$,
 $G^0$ is the goldstone mode, $s$ is the $S$ VEV.  An effective $\mu$ parameter can be defined  
$\mu_{\rm eff} \equiv \lambda s/\sqrt{2}$. As $h_v^0$ is rotated in the same way 
as $v$,
it is the linear combination that gives mass to the $W$ and $Z$  and 
hence is the one that has trilinear couplings to these gauge bosons. Therefore
the Standard Model-like Higgs boson is the one which has the largest component 
in the $h_v^0$ direction, while the nonstandard Higgs boson is the one that is 
mostly in the $H_\nu^0$ direction.

\begin{table}
\centering
\begin{tabular}{|c|c|c|c|c|c|c|c|c|}
\hline
Model & $\lambda$ & $\kappa$ & $t_\beta$ & $A_\lambda$ & $A_\kappa$ & $A_t$& 
$\mu_{\rm eff}$ & $M_{\tilde q}$ \\
 & & & & (GeV) & (GeV) & (TeV) & (GeV) & (TeV) \\
\hline
BM1 & 0.71 & 1.10 & 1.5 & -11.0 & -8.0 & 0.0 &160 &  0.5 \\
BM2 & 0.71 & 1.10 & 1.5 & -9.1 & -7.0 & 0.0 & 166 & 0.5 \\  
BM3 & 0.67 & 0.78 & 1.5 & -4.2 & -40.6 & 0.0 & 170 & 0.5\\  
\hline
\end{tabular}
\caption{Model parameters of benchmark scenarios for enhanced $H_2^0 \to ZA_1^0$
signals.}
\label{tab:bmpara}
\end{table}

\begin{table}
\centering
\begin{tabular}{|c|c|c|c|c|c|c|c|c|}
\hline
Model & $m_{H_1^0}$ & $m_{H_2^0}$ & $m_{H_3^0}$ & $m_{A_1^0}$ & $m_{A_2^0}$ & $m_{H^\pm}$
& $g_{t\bar t H_1^0}^{\rm red.}$ & $g_{t\bar t H_2^0}^{\rm red.}$ \\ 
      & (GeV)      & (GeV)     & (GeV)     &   (GeV)    & (GeV)     & (GeV) & &
\\
\hline
BM1 & 125.2 & 270 & 495 &   8.9 & 357 & 266 & 0.982 &  -0.691\\  
BM2 & 125.1 & 283 & 513 & 19.7 & 365 & 278 & 0.984 & -0.690 \\  
BM3 & 124.5 & 252 & 391 &  117 & 328 & 248 & 0.992 & -0.668\\  
\hline
\end{tabular}
\caption{Higgs mass spectra and normalized coupling to top quark in the benchmark scenarios.}
\label{tab:bmspec}
\end{table}

\begin{table}
\centering
\begin{tabular}{|c|c|c|c|c|c|}
\hline
$\mathcal{BR}$ of $H_1^0$ & $b\bar{b}$ & $ \gamma \gamma$ & $  W W^*$ & $  Z Z^*$ & $  A_1^0 A_1^0$  \\ 
\hline
BM1 & 0.63 & $2.6 \times 10^{-3}$ & $0.19$ & $2.1 \times 10^{-2}$ & $2.9 \times 10^{-3}$ \\
BM2 & 0.61 & $2.5 \times 10^{-3}$ & $0.18$ & $2.0 \times 10^{-2}$ & $4.3 \times 10^{-2}$ \\  
BM3 & 0.64 & $2.7 \times 10^{-3}$ & $0.18$ & $2.0 \times 10^{-2}$ & 0.0 \\
\hline
\end{tabular}
\caption{Relevant branching ratios of the Standard Model-like Higgs boson in the benchmark scenarios.}
\label{tab:h1br}
\end{table}

\begin{table}
\centering
\begin{tabular}{|c|c|c|c|c|c|c|c|c|c|}
\hline
$\mathcal{BR}$ of $H_2^0$ & $b\bar{b}$ & $W W$ & $  Z Z$ & $  H_1^0 H_1^0$ & $  Z A_1^0$ & 
$  A_1^0 A_1^0$ \\ 
\hline
BM1 & $4.5 \times 10^{-3}$ & $1.7 \times 10^{-3}$ & $7.3 \times 10^{-3}$ & $5.6\times 10^{-4}$ & $0.78$ & $0.17$  \\
BM2 & $4.3 \times 10^{-3}$ & $1.6 \times 10^{-3}$ & $7.0 \times 10^{-4}$ & $4.9\times 10^{-4}$ & $0.70$ & $0.16$  \\
BM3 & $1.9\times 10^{-2}$& $1.2 \times 10^{-3}$ & $5.0 \times 10^{-4}$ & $1.7 \times 10^{-6}$ & $0.78$ & $0.19$ \\
\hline
\end{tabular}
\caption{Relevant branching ratios of the lightest non-Standard CP even Higgs boson
in the benchmark scenarios.}
\label{tab:h2br}
\end{table}

\begin{table}
\centering
\begin{tabular}{|c|c|c|c|c|c|c|c|c|c|}
\hline
$\mathcal{BR}$ of $A_1^0$  &$  \tau \tau$& $b\bar{b}$ & $gg$ &Signal Rate ($\mu$) \\ \hline 
\hline
BM1 & $0.74$ & $0.0$ & $0.12$ & 0.28 \\
BM2 & $5.9 \times 10^{-2}$ & $0.92$ & $1.1 \times 10^{-2}$ & $3.7 \times 10^{-3}$ \\  
BM3 & $9.1 \times 10^{-2}$ & $0.87$& $2.9 \times 10^{-2}$ & 0.01\\
\hline
\end{tabular}
\caption{Relevant branching ratios of the lightest CP-odd Higgs boson
in the benchmark scenarios and the signal rate $\mu = \frac{\sigma(pp \to H_2 \to Z A\to Z \tau^+ \tau^-)}{\sigma(pp \to H_{SM})}$ normalized to the Standard Model Higgs cross section.}
\label{tab:A1br}
\end{table}

We consider benchmark points in which the pseudo-scalar is an R-axion
and use NMSSMtools 3.2.3~\cite{Ellwanger:2004xm} to find the
benchmark scenarios with NMSSM parameters shown in Tab.~\ref{tab:bmpara}
and physical Higgs boson masses in Tab.~\ref{tab:bmspec}.  In Appendix~\ref{raxion} we
provide a phenomenological explanation of how such a region of parameter space arises in the NMSSM. As can be seen in the last column of Tab.~\ref{tab:bmspec}, the $H_1 (H_2)$ has production cross sections through gluon fusion $\sim 1 (1/2)$ times the cross section for a Standard Model Higgs at that mass.  The branching ratios are shown in Tab.~\ref{tab:h1br}, Tab.~\ref{tab:h2br} and Tab.~\ref{tab:A1br}. Tab.~\ref{tab:bmspec} and Tab.~\ref{tab:h1br} show that the $H_1^0$ has Standard Model-like   production cross-sections and branching ratios (with deviations ranging 10-30\%) which are consistent with the current sensitivities of the LHC Higgs analyses \cite{ATLASHiggs, CMSHiggs}. To suppress $H_2^0$ couplings to down-type fermions we have chosen to only consider $\tan \beta = 1.5$ scenarios, which also leads to an enhancement in the top quark coupling. The values of $\lambda$ and $\kappa$ are large
in order to generate a significant mass splitting between the two pseudo-scalar states. Some
of these couplings are large enough to develop a Landau-pole before the 
GUT-scale. Hence the UV-completion of such scenarios may require Fat-Higgs
like models discussed in Ref.~\cite{Harnik:2003rs}. For these regions
of parameter space, the Standard Model-like Higgs boson is the $H_1^0$ state
and for $\lambda \lsim 1$, its tree-level mass $m_{H_1^0}^{\rm tree} \lsim 100
$~GeV. Hence  a small amount of stop radiative contributions is needed to raise
the physical SM-like Higgs to the observed Higgs boson mass~\cite{Haber:1990aw, Okada:1990vk, Ellis:1990nz, Barbieri:1990ja}. 

Benchmark point
BM1 was chosen so that $A_1^0$ decays mostly to $\tau$'s due to the phase space
suppression of $A_1^0$ decays to bottom quarks. In the R-axion limit
$m_{A_1^0} \propto \sqrt{s}$, so we raised the mass of $A_1$ by increasing $s$ which
leads to  BM2. In BM2, $A_1^0$ the branching
ratio to bottom quarks is 0.9 while that to $\tau$-leptons is 0.06 because 
$m_{A_1^0} \gg 2 m_b$. Finally, the benchmark point 
BM3 was chosen so to illustrate that regions of NMSSM parameter space exist
where $A_1$ need not be  light and the branching ratio of $H_2^0 \to Z A_1^0$
can still be enhanced.  These three specific benchmark scenarios serve as an important existence proof that it is possible to have significantly 
enhanced  decay rates of $H_2^0 \to Z A_1^0$ compared to the $H_2^0 \to b \bar b $
and still have a $H_1^0$ state with similar branching ratios as a 
Standard Model Higgs.  However, it is important to keep in mind that even more optimistic scenarios are possible for the signal rate.  For example, the $A$ decays into taus could be enhanced in all parts of parameter space in a leptophilic two Higgs doublet model.  

\section{Limits on the simplified model from current searches}
\label{sec:limits}

In this section we take the benchmark points shown in Sec.~\ref{sec:model} to be 
indicative of a generalized model where a nonstandard Higgs boson $H$ 
dominantly decays into the $Z$ boson and a lighter pseudo-scalar $A$.  From now on, we proceed model-independently and analyze the phenomenology of the signal process $H\to Z A \to Z \tau^+ \tau^-$ in the multilepton decay channel for a broad range of $H, A$ masses.     
We start by finding the efficiency of observing this model in the CMS multilepton analyses under the selection cuts in Ref.~\cite{CMS2}. 


\subsubsection*{Event simulation}
We generate samples of signal events for a broad range of $H, A$ masses using 
Pythia8.170~\cite{Sjostrand:2007gs} including the effects of initial state 
radiation, final state radiation, multiple interactions and fragmentation. 
These events samples were generated for $pp$ collisions at $\sqrt{s} = 7$~TeV
using the CTEQ5L parton distribution functions~\cite{Pumplin:2002vw} 
events we only consider the leptonic decay of the Z-boson (including $\tau$'s) and assume that the
$A$ decays only into $\tau$-leptons.  We do not apply any detector simulation on the events, in order to apply our own tau reconstruction.

\begin{figure}
\begin{center}
\includegraphics[width=.6\textwidth]{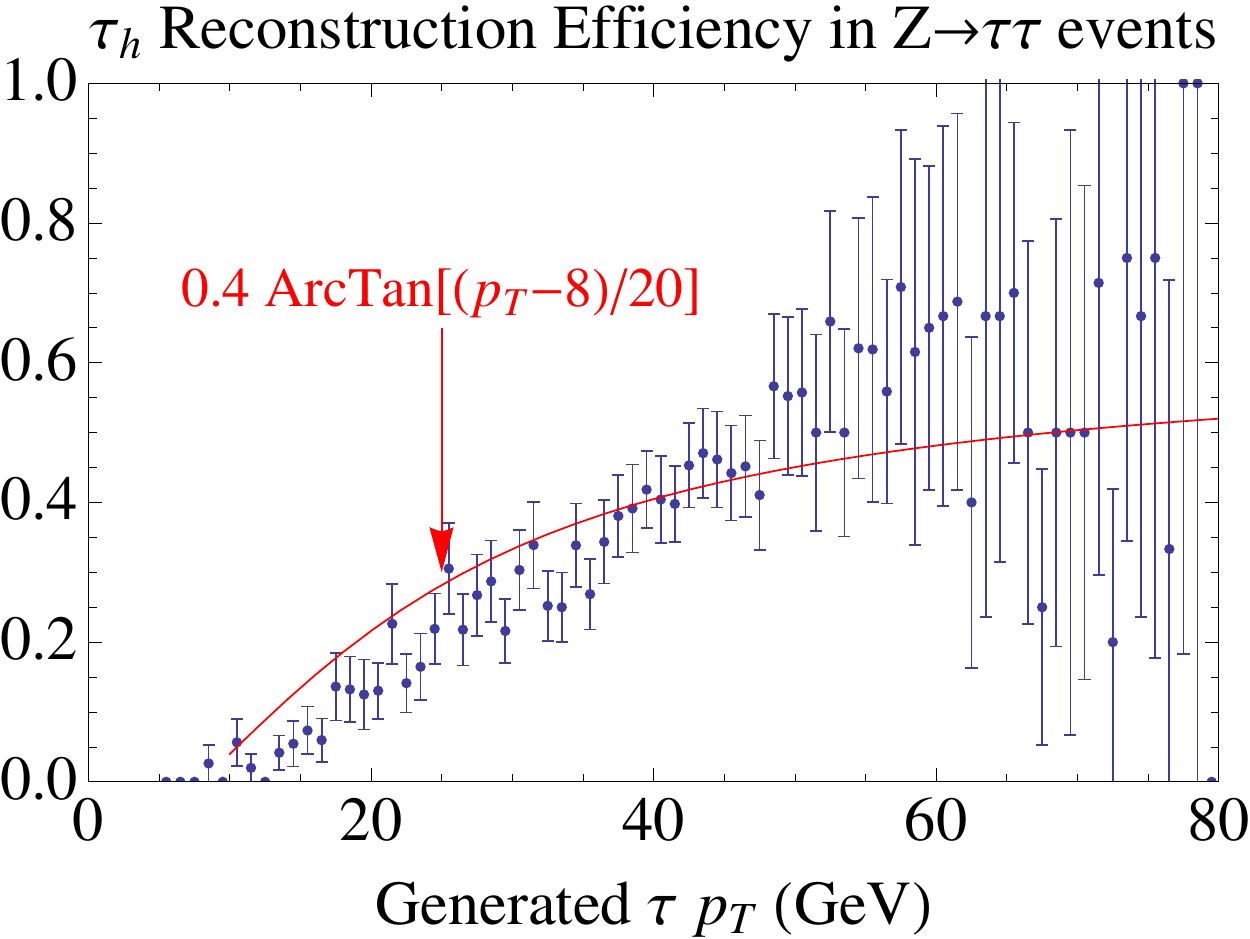}
\end{center}
\caption{Our $\tau$ reconstruction efficiency for a sample of Drell-Yan $Z\to \tau \tau$ events where one tau decays in a one-prong hadronic decay as a function of the tau's generated $p_T$.   The efficiency asymptotes to $\sim 60$\% at high $p_T$ and we have plotted an analytic function to guide the eye. }
\label{fig:Ztaueff}
\end{figure}

The selection criteria for the CMS multilepton analysis of Ref.~\cite{CMS2} are as follows.
The two trigger requirements were ---  {\em single lepton:} one muon (electron) has a $p_T > 35\, (85)$~GeV and {\em dilepton:}  two of the 
leptons have $p_T^1 \geq 20$~GeV and $p_T^2 \geq 10$~GeV.  An electron or muon is identified, if its $p_T \geq 8
$~GeV and $|\eta| \leq 2.1$. Furthermore an isolation criterion 
$I_{\rm Rel} = E_{\rm cone}/E_l \leq 0.15$ is imposed on the leptons, where $E_l$ is the 
energy of the lepton and $E_{\rm cone}$ is the surrounding visible energy in a 
cone of $\Delta R = 0.3\,(0.4)$ for muons (electrons). For reconstructed hadronic $\tau$-leptons we 
consider only one-pronged hadronic decays where the $p_T$ of the charged hadron
is required to be $\geq 8.0$~GeV.  A  tau is isolated if $I_{\rm Rel} \leq 0.15
$ where $I_{\rm Rel}$ is the ratio of total energy inside an annulus with $
0.1 < \Delta R \leq 0.3$ to the total energy inside a cone with $\Delta R 
\leq 0.1$.  To speed up the analysis, we place a lower cut on tracks of $p_T > 0.3$ GeV to be considered in the analysis and have checked that tau reconstruction efficiencies are insensitive to this cut.

Ref.~\cite{CMS2} uses the CMS particle 
flow algorithm to identify the neutral pions and $p_T$ of the $\tau_h$
candidate, which may lead to a
difference between our simulation and the CMS data.  As a check of our tau reconstruction, we simulated one-prong tau decays in Drell-Yan $Z\to \tau \tau$ and found reconstruction efficiencies as shown in Fig.~\ref{fig:Ztaueff} as a function of the generated tau $p_T$.  Our asymptotic efficiency is about 60\% at high $p_T$.  This can be compared with the published CMS tau efficiencies in figure 3 of Ref.~\cite{CMS-PAS-TAU-11-001}.  There is a difference in presentation since the CMS figures are plotted with respected to the generated visible tau $p_T$ which is approximately $1/2-1/3$ of the generated tau $p_T$.  However, taking this into account,  the behavior is similar to the TANC medium tau algorithm and systematically higher than the efficiencies of the HPS algorithms (which peak at 50\%).  This gives us confidence that our tau reconstruction is realistic and maybe just a bit more optimistic than the true CMS algorithms.

 As the $Z$-boson can also decay to $\tau$-leptons which
can further decay into $e$ or $\mu$, 
there is also a possibility of this process also contributing to the multilepton
events in Ref.~\cite{CMS1} and Ref.~\cite{CMS2} where the invariant 
mass of no two leptons are within the $Z$ mass window $75 \; \text{GeV}\leq m_{\ell^+ \ell^-} \leq 105 \; \text{GeV}$. However these 
electrons and muons are typically soft and therefore such events have more difficulty with the triggering requirement.

\begin{figure}
\begin{center}
\includegraphics[width=.32\textwidth]{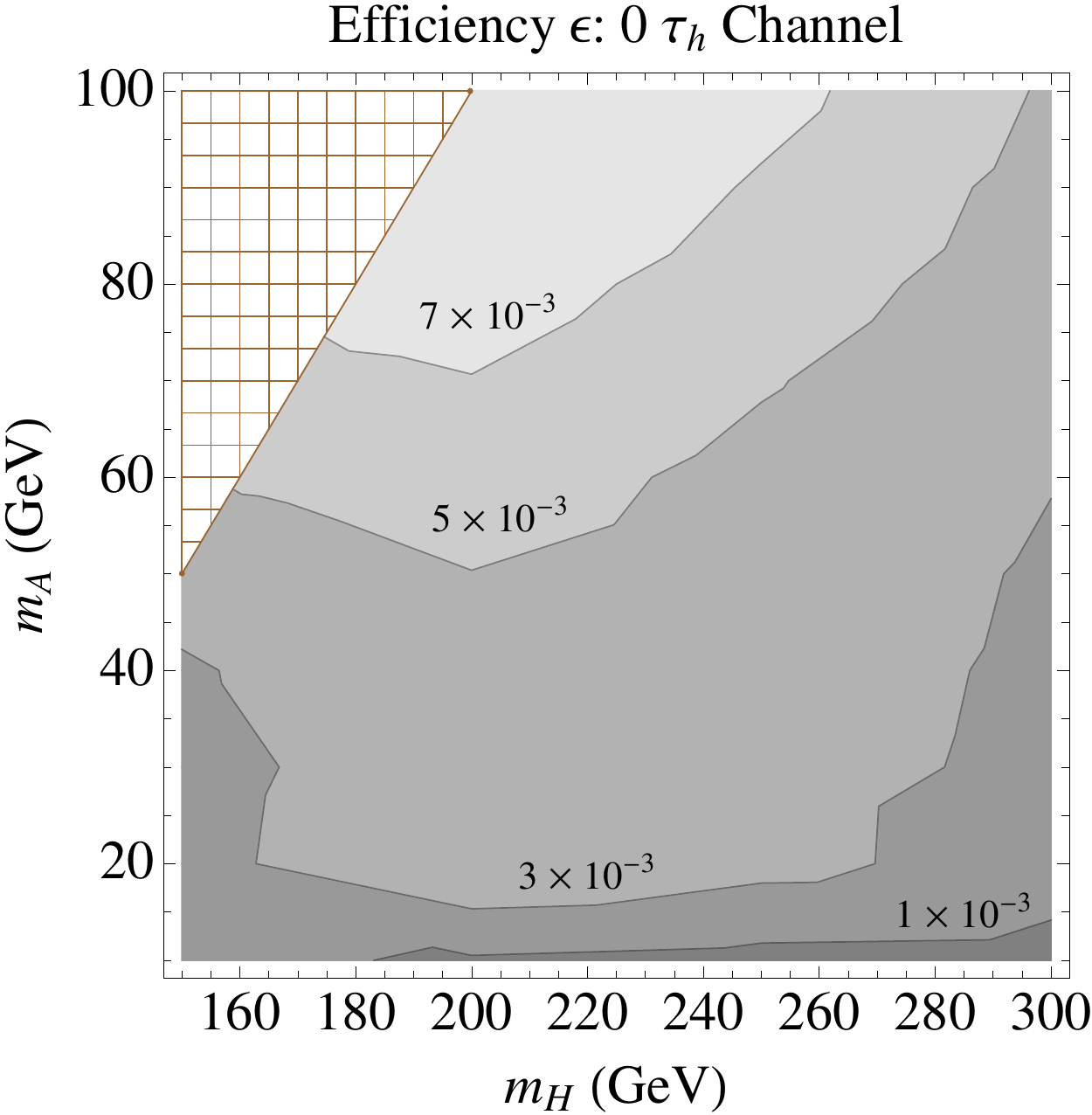}
\includegraphics[width=.32\textwidth]{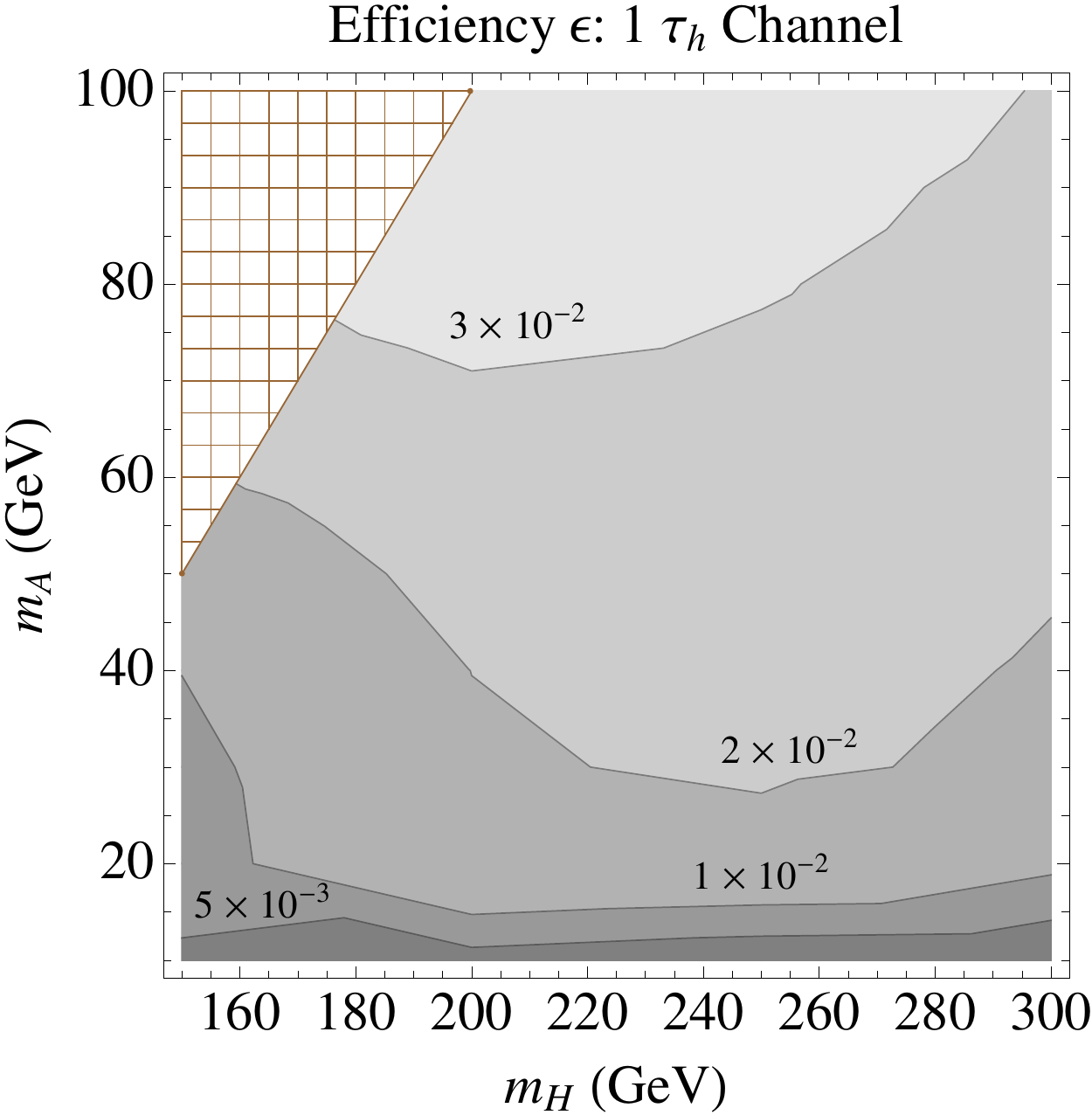}
\includegraphics[width=.32\textwidth]{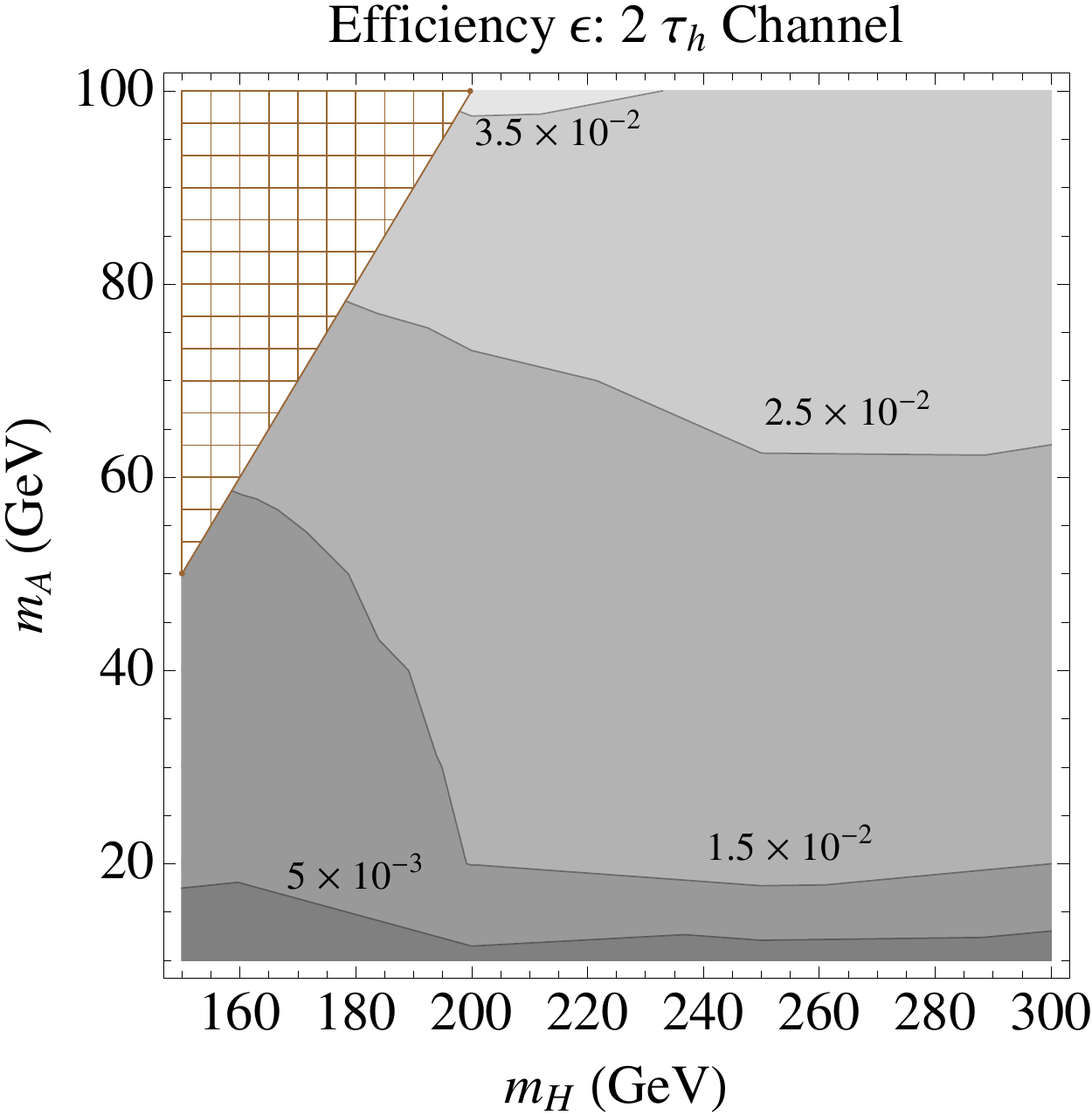}
\end{center}
\caption{Contours of the efficiencies in the $0$ one-prong hadronically decay $\tau$ channel, $1$ one-prong 
hadronically decay $\tau$ channel and $2$ one-prong hadronically decay $\tau$ channel.  In these channels, there is an electron or muon pair with mass consistent with the $Z$ and a total of four leptons (electrons, muons, and one-prong taus).  The brown hatched regions in the upper left are plotted to avoid the kinematically squeezed region where $m_H  - m_A < 100$ GeV. 
}
\label{fig:taueff}
\end{figure}

In Fig.~\ref{fig:taueff} we show the contours of the efficiencies 
for each channel, 
\bea
\epsilon_i \equiv \frac{N(selected)_i}{N({Z\to (e^+ e^-,\mu^+\mu^-, \tau^+\tau^-), A\to \tau^+ \tau^-})}
\eea
where the channels are the 4$\ell+0\tau_h$, 3$\ell+1\tau_h$ and $2\ell+2\tau_h$ channels with a reconstructed leptonic $Z$.  Thus, our convention takes out the  $Z$ branching ratio into the three generations of leptons and the $A$ branching ratio into $\tau$ pairs, but includes the $\tau$ branching ratios into the efficiency.  
The brown hatched regions in the upper left are plotted to avoid the kinematically squeezed region where $m_H  - m_A < 100$ GeV. 
In the $0 \tau_h$ channel, each of the $\tau$-leptons from the $A$ decay have decayed leptonically
via the three-body decay. Thus,  the resulting leptons from the 
decay of the $\tau$'s are relatively soft. This explains the efficiency improvement as $m_A$ increases, since the $\tau$'s have larger average $p_T$ 
and are more geometrically leading to better isolation.   The $1\tau_h$ and $2\tau_h$
efficiency curves have a similar structure to the $0\tau_h$ events 
because of the similar effects of boosting the $A$ and $\tau$-leptons. 
However, the value of $m_H$ with optimal efficiency for the $1\tau_h$ and  
$2\tau_h$ channels is different because of the different isolation requirements
for 1-pronged hadronic $\tau$'s and the greater visible $p_T$ in such decays
as compared to the leptonic case.   This helps to explain the slope of the contours, as a smaller boost to the $A$ is required for hadronic taus to pass the $p_T$ selection.

\begin{figure}
\begin{center}
\includegraphics[width=.45\textwidth]{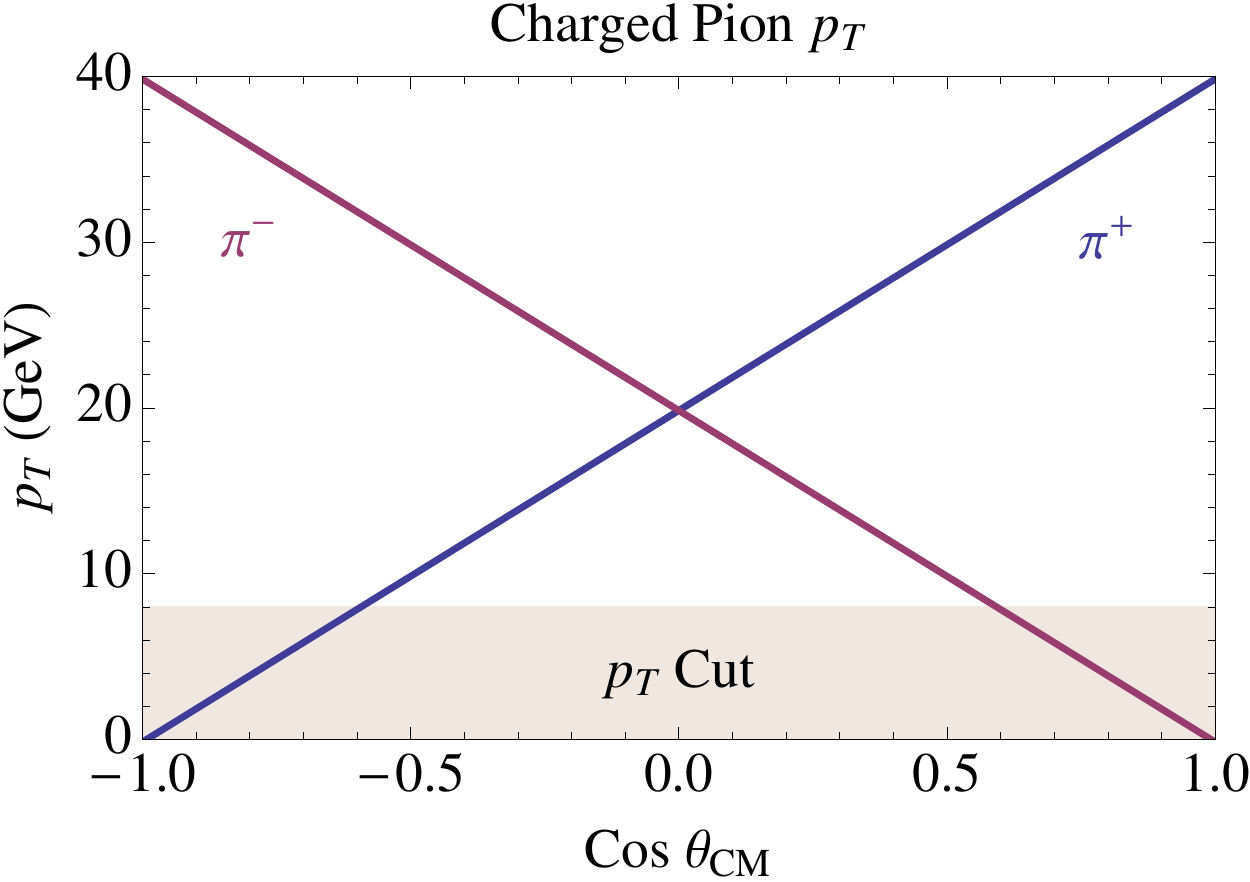} \hspace{.2in}
\includegraphics[width=.43\textwidth]{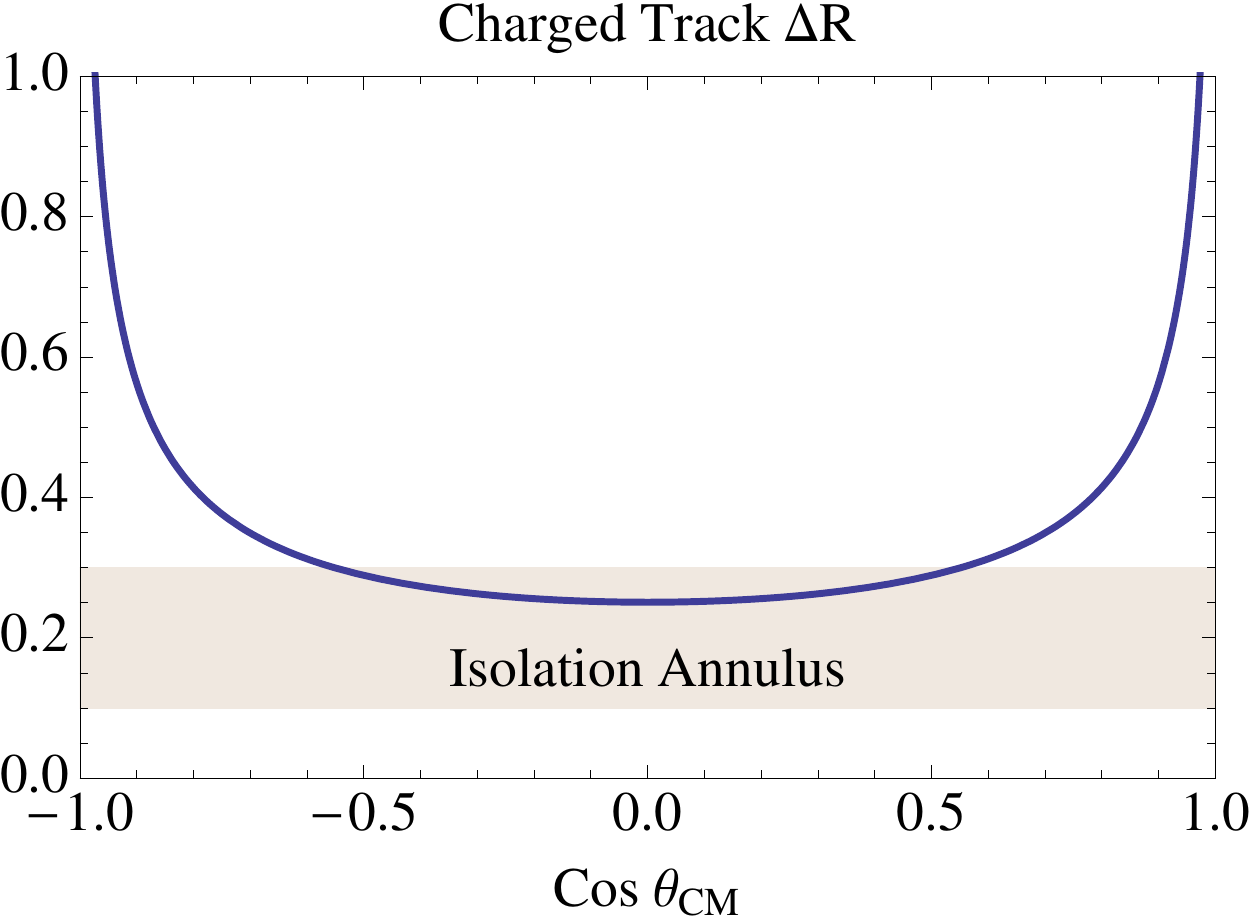}
\end{center}
\caption{Toy model kinematics of the charged pions as a function of the decay angle of the $\tau^+$ in the rest frame of $A$.  The $p_T$ selection cut is shaded on the left plot and the isolation annulus is shaded on the right plot.  The masses for these plots are $m_H, m_A = 200, 10$ GeV.}
\label{fig:toymodel}
\end{figure}

Using a toy model, we can get further insight into the inefficiency of the searches at low $m_A$.  In our toy model, we assume that $H$ is produced at rest, with the $A$ particle being produced in the transverse direction.  This $A$ is taken to decay into two single prong $\tau$'s where the $\tau$'s decay $\tau^+ \to \pi^+, \bar{\nu}_\tau$, assuming $p_{\pi^+} = p_{\tau^+}/2$.  In this case, we can very simply predict the visible pion kinematics as a function of the $\tau^+$ decay angle in the $A$ rest frame.  The $p_T$ and $\Delta R$ of the charged pions are shown in Fig.~\ref{fig:toymodel} for the case of $m_H, m_A = 200, 10$ GeV.  Here, we see that the tau reconstruction requirement of a track with $p_T > 8$ GeV restricts us to  $|\cos \theta_{CM}| \lesssim .6$  in order to select both hadronic $\tau$'s.  However, as the $\Delta R$ figure shows, the isolation condition is in direct conflict,  requiring $|\cos \theta_{CM}| \gtrsim .6$.  Thus, these two conditions are in tension.  Due to the softness of the pions, the configuration that works best for getting substantial $p_T$ is where the two taus decay in the longitudinal direction, so that the boost enhances both of their transverse momenta.  However, at the same time this pushes the pions on top of each other, worsening isolation.  This tension is exacerbated with larger boosts.  For example, as the $H$ mass is increased, the slope of the $p_T$ plots increases whereas the dip of the $\Delta R$ plot decreases.  Thus, we see that the standard tau selection and reconstruction is inefficient for the boosted regime.  For such decays, searches for boosted taus has been show to be efficient  \cite{Englert:2011iz,Katz:2010iq}, in particular using $N$-subjettiness \cite{Thaler:2010tr}.  Thus,  multilepton analyses should consider a boosted tau pair object as a way to recover such regions of parameter space.

\subsubsection*{Signal limits using the CMS analysis}
The multilepton analysis Ref.~\cite{CMS2} observed number of events and expected background is tabulated in Tab.~\ref{tab:mhmalim}. 
There is a slight excess, predominantly in the $2\tau_h$ channel, which unfortunately is not fit well by our signal.
This is due to the fact that the efficiencies for the $1\tau_h, 2\tau_h$  channels are similar (see Fig.~\ref{fig:taueff}), which limits the amount of the excess that can be explained.    
In order to set limits on our model, we use this data to calculate the maximum number of signal events
at 95\% C.L.~ in each channel. 
Due to 
the low statistics we assume a Poisson distribution for 
the number of events and therefore the probability of observing at least $N_{\rm obs}$ 
events due to signal and background is $P(N_{\rm obs}|S+B) = \Gamma(N_{\rm obs}+1,S+B)/N_{\rm obs}!$, 
where $\Gamma(A,B)$ is the incomplete gamma 
function. We assume the background has a gaussian distribution
with mean $\mu_B$ and variance $\sigma_B^2$ and find the maximum allowed number
of signal events at 95\% C.L. ($S_{95}^{\rm Max}$) by solving the equation
\bea
\int_0^\infty dB \frac{\Gamma(N_{\rm obs}+1,S_{95}^{\rm Max}+B)}{N_{\rm obs}!} \frac{1}{\mathcal{N}_B} 
\exp\left[-\frac{(B-\mu_B)^2}{2\sigma_B^2} \right]  = 0.05
\eea
where $\mathcal{N}_B$ normalizes the background's gaussian distribution over the interval $[0,\infty)$.
\begin{table}
\centering
\begin{tabular}{|c|c|c|c|c|c|}
\hline
Selection,     & Obs. Events &  Expected Bkg & Bkg Err.          & Max. All. & Projected \\ 
Channel        & $N_{\rm obs}$       &  $\mu_B$   & $\sigma_B$ & $S_{95}^{\rm Max}$ & $S_{95}^{\rm 30 fb^{-1} Max}$ \\
\hline
$0 \tau_h$ &      33       &    37      &    15             & 22.6 & 36.4 \\
 $1 \tau_h$ &      20       &    17      &    5.2            & 15.6 & 25.0\\
  $2 \tau_h$ &      62       &    43      &    16             & 48.9 & 38.7 \\
\hline
\end{tabular}
\caption{The observed number of events and expected backgrounds for the three channels as given in Ref.~\cite{CMS2}.  In the second to last column is our derived $95\%$ C.L.~limit  on the number of signal events in each channel and the last column is the projected limit a 30 fb$^{-1}$ analysis would have given only statistical background errors.}
\label{tab:mhmalim}
\end{table}
Using these limits on the number of signal events in each channel for each selection,
we can put bounds on the cross-section for this process, normalized to the Standard Model Higgs 
production cross-section by defining a signal strength parameter
\be
\mu_{95}^i \equiv \frac{S_{95}^{i\,{\rm Max}}}{\sigma_{\rm H_{SM}} \times \mathcal{BR}(Z \to l^+l^-) \times \epsilon^i
\times \mathcal{L}}
\ee
where $l=e,\mu,\tau$ and $\epsilon^i$ is the efficiency for i$^{th}$ channel and $\mathcal{L}$ is the integrated luminosity.  

\begin{figure}
\begin{center}
\includegraphics[width=.4\textwidth]{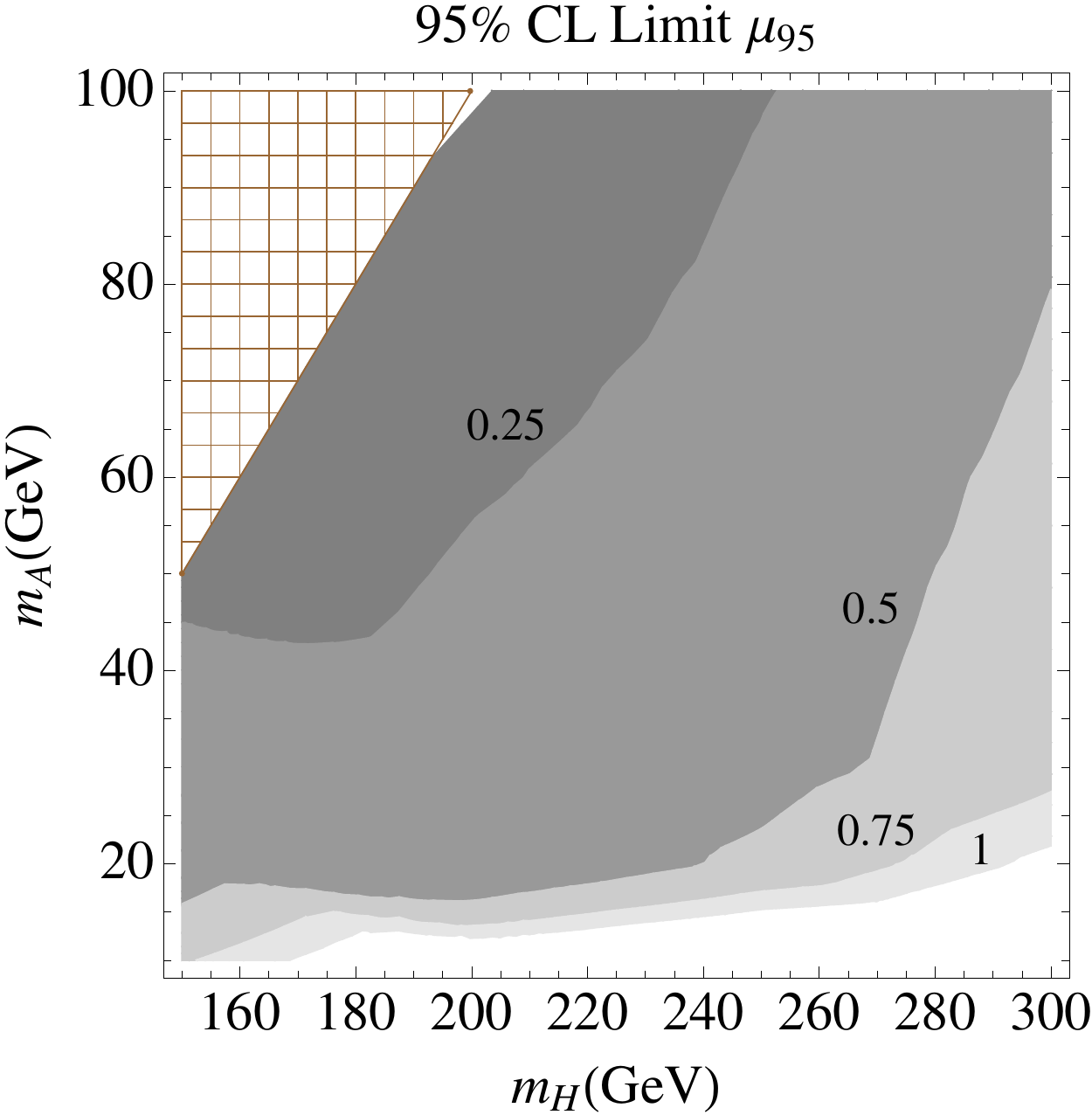} \quad \includegraphics[width=.4\textwidth]{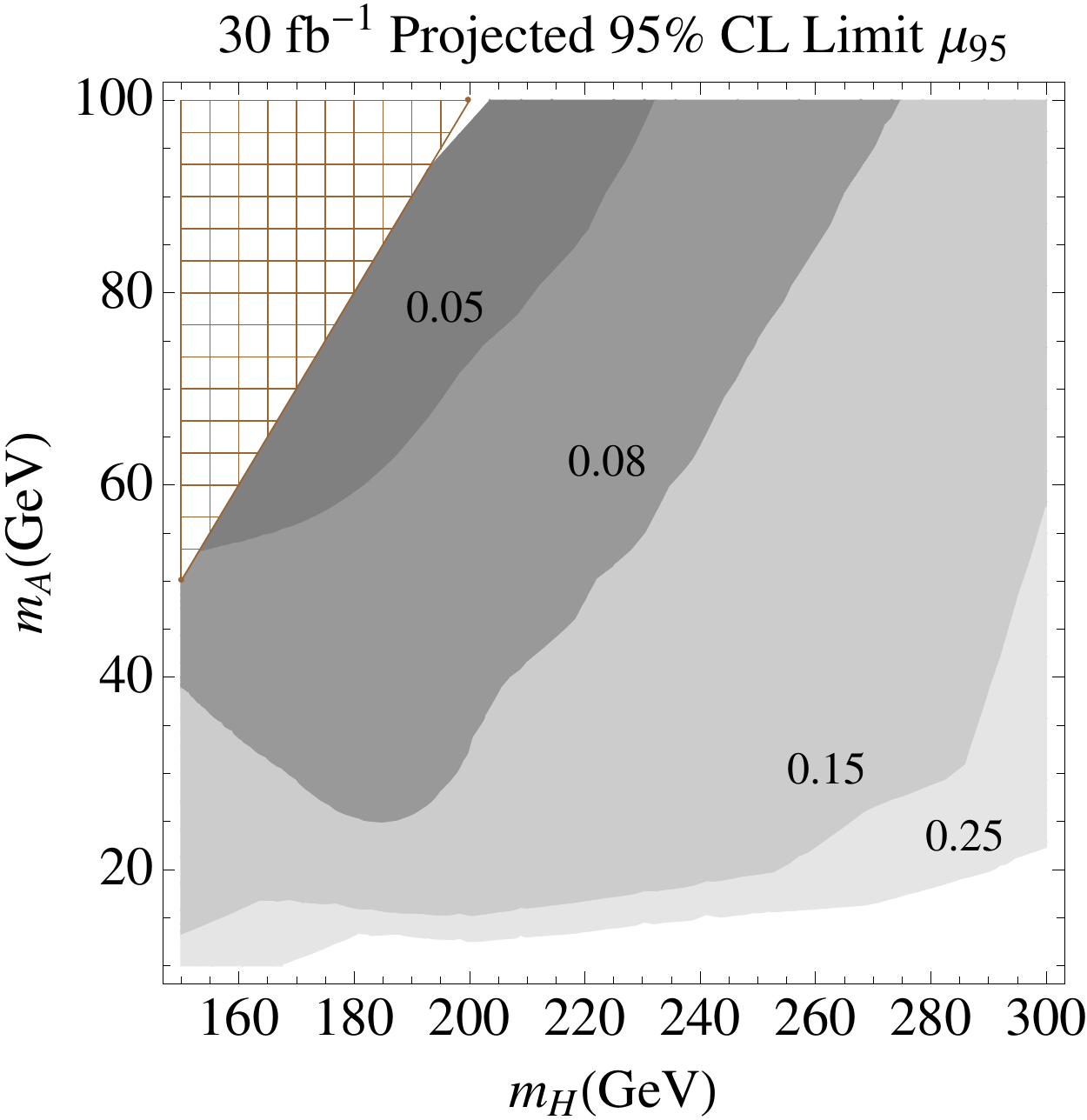}
\end{center}
\caption{In the left figure, the contours of the 95\% CL limit on the signal parameter $\mu_{95}$ is displayed, while in the right figure, the projected limit a 30 fb$^{-1}$ CMS-like analysis would have.}
\label{fig:mumin}
\end{figure}

In Fig.~\ref{fig:mumin} we present contours of the minimum value of $\mu_{95}$ for
the $\mathcal{L } = 4.8$~fb$^{-1}$  of data. 
The limits are weaker for small values of $m_A$ because of 
the lower efficiencies and they are stronger for small values of $m_H$ because of
the larger production cross-section.
Using the couplings in Tab.~\ref{tab:bmspec} and the branching 
ratios in Tab.~\ref{tab:bmbr} we see that these benchmarks have 
$\mu \sim 0.25$ for 
$m_A = 10$~GeV and $\mu \sim 
0.023$ for $m_A \gsim 10$~GeV. Therefore the benchmark points considered 
in Tab.~\ref{tab:bmpara} are not constrained by the present 
experimental data.  However, in more optimistic scenarios where the $A$ is leptophilic (i.e. $Br(A_1^0 \to \tau \tau) \sim 1$), we see that  the Standard Model cross section is already strongly constrained for $A$ masses above about 15-20 GeV. 

With improvements in statistics and in the background modeling, a future multilepton analysis would have improved sensitivities to this signal and start to constrain more interesting signal rates.  To estimate this improvement, in Tab.~\ref{tab:mhmalim}, we have projected the signal events allowed in a 30 fb$^{-1}$ CMS-like analysis.  In making this estimate, we have ignored the background and signal cross section changes with  $\sqrt{s}=8$ TeV running and further assumed that the background uncertainty can be reduced to being purely statistical.  As shown in the right figure of Fig.~\ref{fig:mumin}, this gives a projected limit at 30 fb$^{-1}$ that is roughly $4-5$ times stronger than the current analysis.   This is very close to being sensitive to our benchmark signal rates and would place stringent constraints on scenarios where the $A$ branching ratio to taus is enhanced.  
Hence, a multilepton analysis using the full LHC data set of 2012 could have an interesting reach for for this nonstandard Higgs signal.        

\section{Comparison of methods of mass reconstruction $H \to ZA$}
\label{sec:recon_comp}

If an excess in these channels is seen in future multilepton analyses, it will be important to reconstruct the signal in order to determine the underlying theory.  
As a step in this direction, in this section we consider reconstructing this signal by 
measuring the masses $m_H$ and $m_A$ through a variety of techniques. In particular
we consider three possibilities:  $i)$ transverse
mass, 
$ii)$  collinear mass  and $iii)$ an analytic solution based on trial masses for $H, A$. In this section we will be 
concentrating purely on the $2\tau_h$ channel because there are more 
neutrino final states in the $0\tau_h$ and $1\tau_h$ channels, complicating the reconstruction.  We summarize these mass reconstruction methods below.

{\em Transverse Mass Variables:  }  Using the visible components of the $\tau$'s $p_{V_{1,2}}$, the
reconstructed $Z$ momentum $p_Z$ and the total MET components $p_+^T$
we can define the transverse masses~\cite{Barr:2009jv}
\bea
m_A^T &=& \sqrt{p_V^2+ 2(E_V E_+^T - p_V^T \cdot p_+^T)} \label{eq:mAT}\\
m_H^T &=& \sqrt{(p_V+p_Z)^2 + 2((E_V+E_Z) E_+^T - (p_V^T + p_Z^T) \cdot p_+^T)}
\label{eq:mHT}
\eea
where $p_V = p_{V_1} + p_{V_2}$.  These variables have the property that $m_{X}^T \leq m_X$, so measuring the endpoints gives a determination of the masses. 

{\em Collinear Approximation:}  Since the pseudo-scalar $A$ typically has a large boost, it is a good approximation
to assume that the final state neutrinos are collinear with the visible
final state hadrons of the taus. In this approximation, the neutrino momenta are
proportional to the visible components of the $\tau$-leptons and we need
to solve the linear equation~\cite{Ellis:1987xu}
\bea
\lambda_1\, p_{V_1}^T + \lambda_2\, p_{V_2}^T = p_+^T.
\eea
Physical solutions require the coefficients $\lambda_{1,2}$ to be positive.  From these, one approximates the $\tau$ momenta as $(1+\lambda_i) p_{V_i}$ to determine the $A, H$ masses.   

{\em Analytic Solution:} Finally, as shown in Appendix \ref{sec:mhmatest}, for trial masses $m_H$ and $m_A$
we can solve the the neutrino momenta exactly.   Similar to Ref.~\cite{Cheng:2007xv}, for each event, there is an allowed region of masses where there are consistent neutrino solutions.  For each such event, our estimator for the masses is the center of mass of the allowed $(m_H, m_A)$ region.
See Appendix \ref{sec:mhmatest}, for more details on this method. 

\begin{figure}
\begin{center}
\includegraphics[width=.45\textwidth]{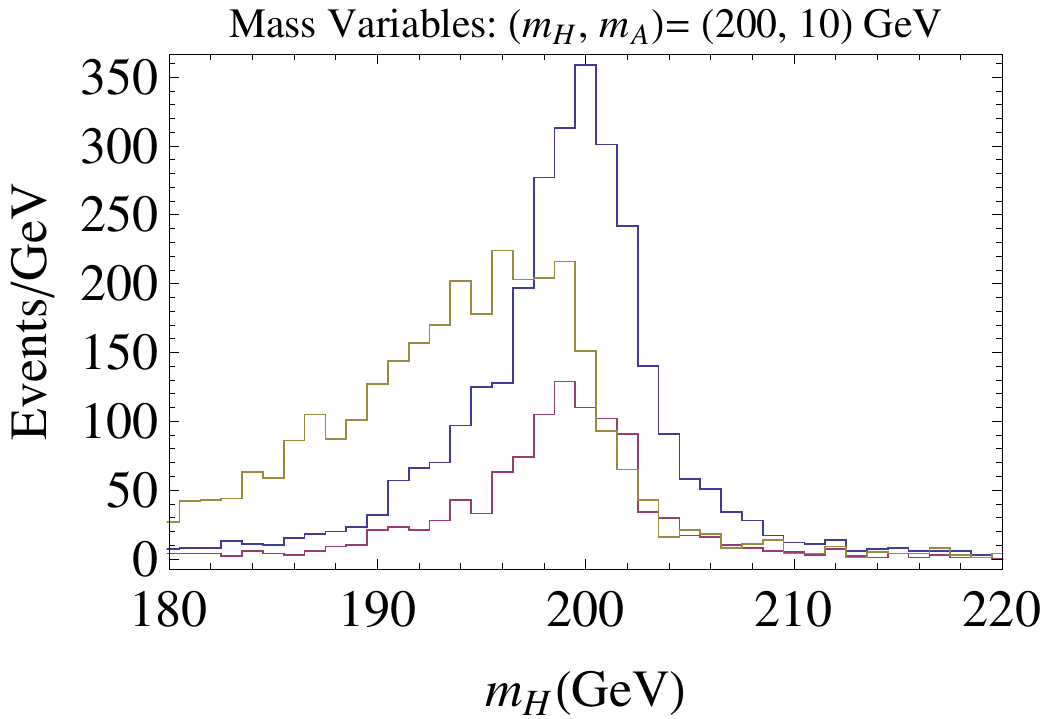}
\includegraphics[width=.45\textwidth]{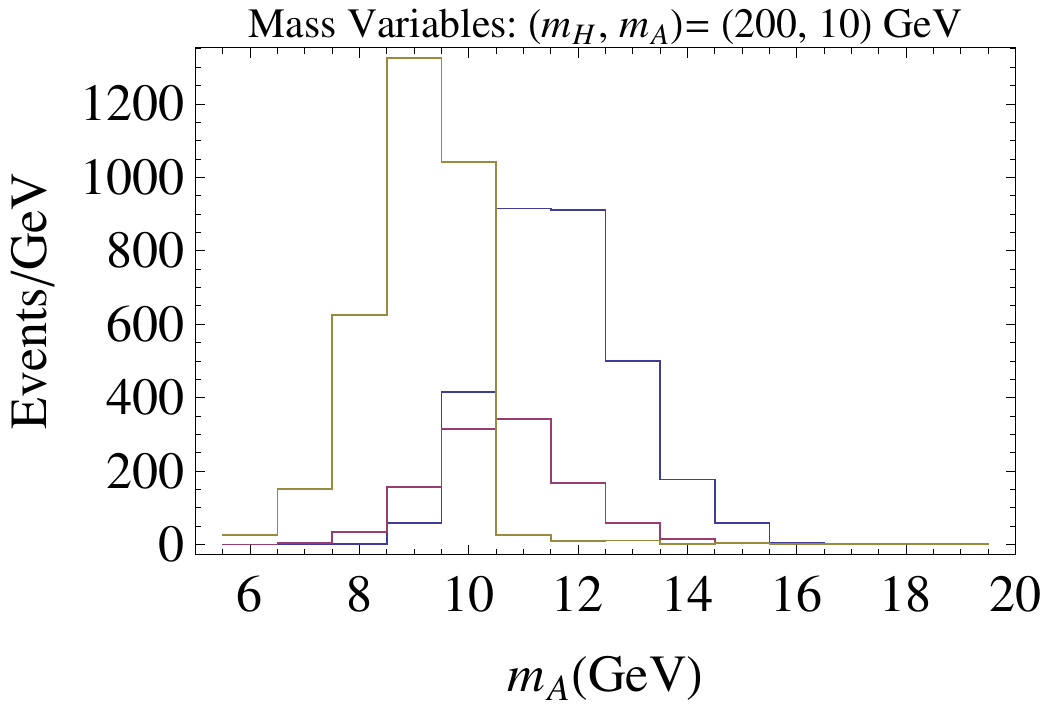}
\includegraphics[width=.45\textwidth]{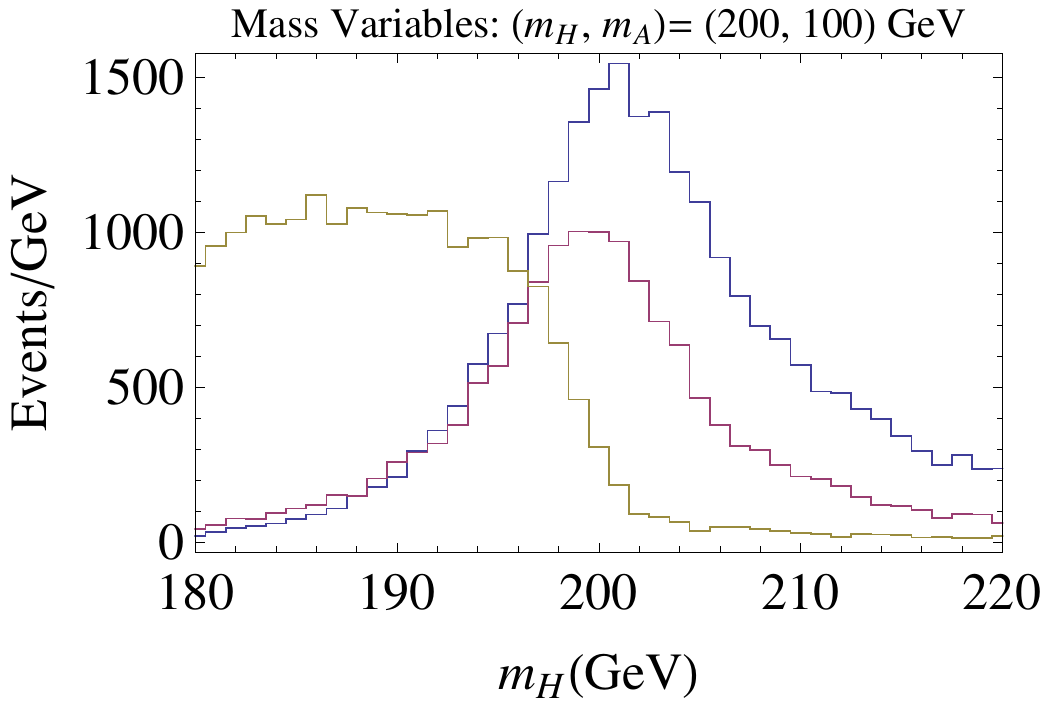}
\includegraphics[width=.45\textwidth]{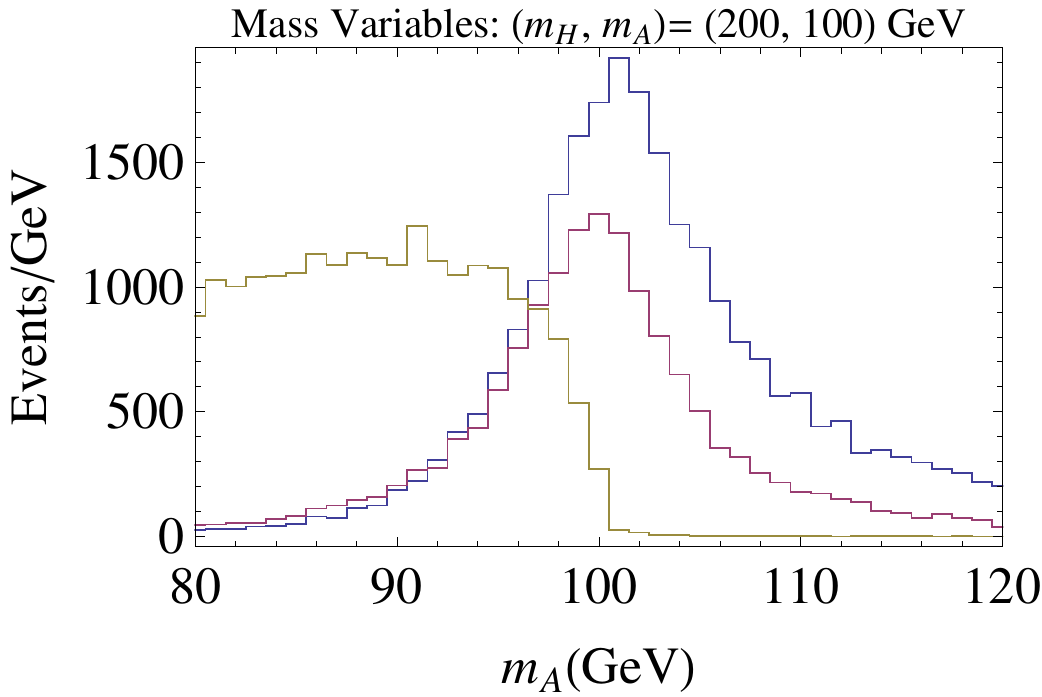}
\includegraphics[width=.45\textwidth]{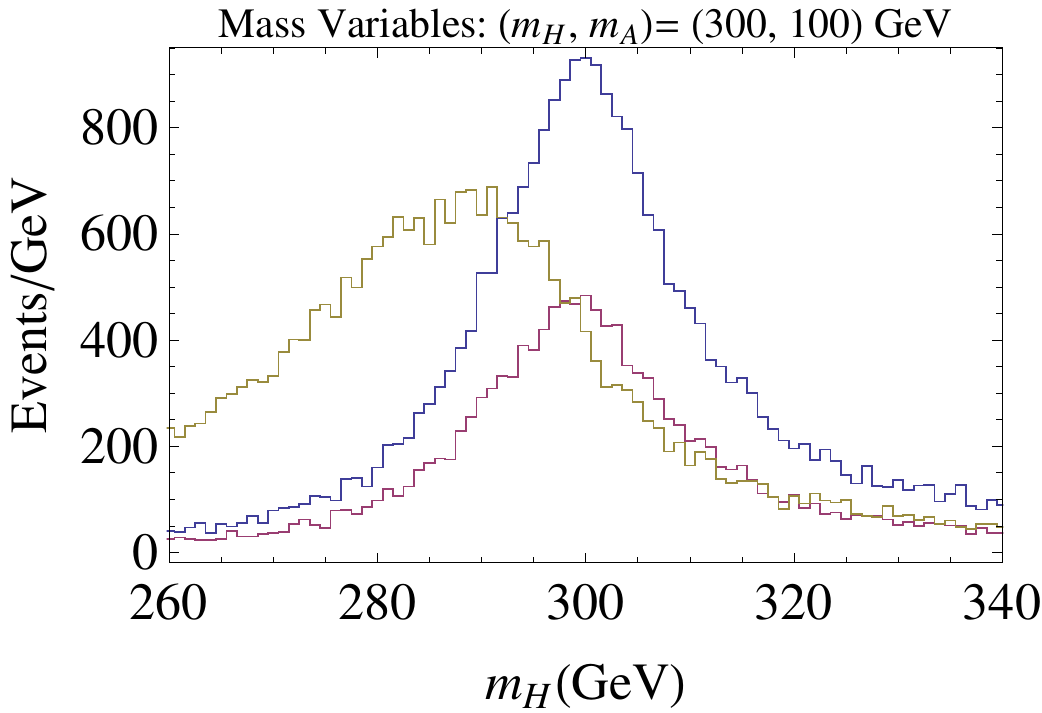}
\includegraphics[width=.45\textwidth]{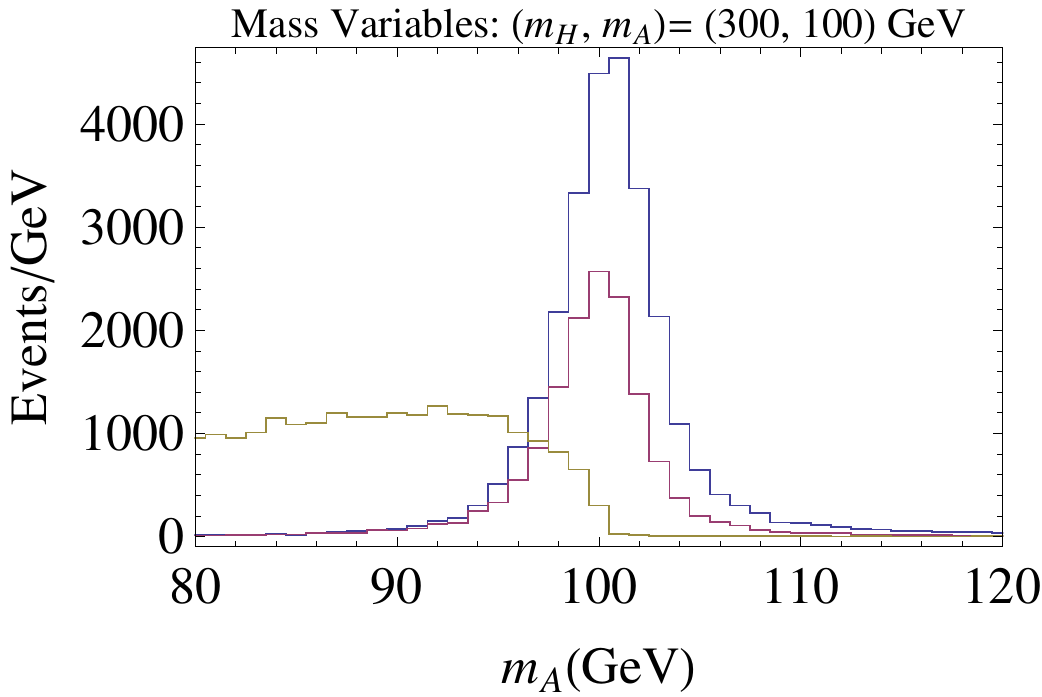}
\end{center}
\caption{Comparison of the reconstruction of mass variables for a $(m_H ,m_A) = (200~\mbox{GeV}, 10~\mbox{GeV}), (200~\mbox{GeV},100~\mbox{GeV})$ and $(300~\mbox{GeV},100~\mbox{GeV})$. 
The left panels are the reconstruction of the $H$ mass while the right panels correspond to those of
the $A$ mass. The yellow (light grey) histogram corresponds to the transverse masses of 
$H$ and $A$ method $i)$ and its drop off determines the mass, the purple (grey) histograms correspond to the mass reconstructed by using the collinear solution to the neutrino momenta method $ii)$ and the blue (dark grey) histograms corresponds to the reconstructed mass using the analytic method $iii)$.}
\label{fig:mhvsma}
\end{figure}


In Fig.~\ref{fig:mhvsma} we show the histograms for each of these different
mass variables for signal events generated with
$(m_H,m_A) = (200~{\rm GeV},10~{\rm GeV}),(200~{\rm GeV},100~{\rm GeV})$ and $(300~{\rm GeV},100~{\rm GeV})$. The blue (darker grey) histograms
correspond to method $iii)$, the average values of $m_H$ and $m_A$ that provide 
analytic neutrino solutions for each event. The yellow (lighter grey) histograms 
correspond to the transverse masses for each event as defined in Eq.~(\ref{eq:mAT}) and 
Eq.~(\ref{eq:mHT}). The purple (grey) histograms are the $m_H$ and $m_A$ masses
calculated in the collinear approximation.  The peaks of the blue and 
purple histograms determine the reconstructed mass of $m_H$ and $m_A$, while the
transverse masses (shown in yellow) give another estimate of the masses as the transverse mass should drop above the physical mass. Hence the peaks for blue and purple histograms are to be compared with the value of the mass where the yellow histograms drops down.  As can be seen by the plots, this drop off is not always sharp, so an accurate extraction of the mass from the  transverse mass distribution can require a good understanding of its shape.  

The average solutions shown by the blue histogram  typically reconstruct an
$m_H$ that agrees well with the true value in the signal, although the value
of $m_A$ is slightly higher than its true value for low values of $m_A$. This slight bias of reconstructed $A$ masses, at low values of $m_A$, is due to the 
selection requirements preferring taus with a higher $p_T$. Furthermore about $90\%$ of the $2\tau_h$ events that pass the
isolation and trigger cuts, reconstruct to a physical solution. This is to be compared to the collinear solution
where less than $50\%$ of the $2\tau_h$ events lead to physical solutions with positive $\lambda_{1,2}$.  The plotted event counts reflects this difference in solution efficiency, giving a visual indication of the increase in reconstructible events.  Thus, for a given set of events, the analytic method can provide a more accurate and efficient measurement of the $H$ and $A$
masses. The reconstructed masses using the collinear method are typically slightly lower than
physical masses because that angles between the neutrinos and visible hadrons are neglected. 
From Fig.~\ref{fig:mhvsma},  we see that the widths of the distributions using methods $ii)$ and $iii)$ are comparable. Comparing the transverse mass distributions for $m_H = 200$~GeV and $m_H = 300$~GeV, we see that the mass resolution of $H$ and $A$ using the end points of the distributions is better for lower masses than for higher masses. 
Therefore, the mass determination using the transverse mass distribution is more complicated for these higher mass points.  

To summarize, we have compared three different mass variables as a means to reconstruct this signal process.  These mass reconstruction methods all work reasonably well, displaying different advantages and disadvantages depending on the situation.  We focused on the  $2\tau_h$ channel, since there the kinematics can be solved once trial masses for $A, H$ are given.  
Even though the analytic method can be quite sensitive to the measured visible momenta, most 
of the $2\tau_h$ solutions lead to a physical solution. This result is to be compared to the collinear method where a majority of the events do not have have a physical solution and the
transverse mass distribution whose endpoint may be uncertain.   
In our reconstruction, the only uncertainty was the decay widths of the heavy particles, thus our analysis gives an idea of the irreducible uncertainties without considering detector effects.  
 To improve further,    there are more advanced techniques being developed which could yield further improvements.  Likelihood methods taking into account the $\tau$ decay kinematics, such as \cite{Elagin:2010aw}, would be worth applying to this process.  

\section{Conclusion}
\label{sec:concl}

In this article we have shown that it is possible for nonstandard CP-even 
Higgs bosons to have large branching fraction into $ZA$ where 
$m_A \lsim 100$~GeV and $A$ has a sizeable branching fraction into 
$\tau$-leptons. In particular, in the NMSSM we have presented three benchmark 
scenarios where the $H_2^0 \to ZA_1^0$ decay mode has the largest branching 
fraction while $H_1^0$ phenomenology is SM-like. In these scenarios, low $\tan \beta$ and large mixing in the pseudo-scalar sector are 
preferred. These rates can be further enhanced in more optimistic scenarios
like in leptophilic two Higgs doublet models.

In the collider study of the $H \to Z A \to Z \tau^+ \tau^-$ scenario we found 
the efficiencies for passing the selection cuts in the  
$0\tau_h$, $1\tau_h$ and $2\tau_h$ channels that included a leptonic $Z$ in the latest CMS multilepton analysis \cite{CMS2}. The shape of the efficiencies 
contours for each channel are due an interplay between the selection cuts,
the isolation requirements and the kinematics of the events.  Using a toy model, we demonstrated that the low $m_A$ parameter space is inefficiently picked up by the multilepton selection.  In particular, the transverse momenta and isolation requirements for the taus are in direct tension with each other.  This motivates including boosted di-tau jets, as explored in \cite{Englert:2011iz}, as a physics object in future multilepton analyses.
Using the observed and expected background events  we constructed the 95\% C.L. limits on the
number of signal events in each channel. We found that the strongest limit on 
such a scenario was due to the $1 \tau_h$ channel because of a high signal 
efficiency and the consistency between expected and observed events.   In addition, we made a projection of the reach of a 30 fb$^{-1}$ CMS-like study would have on this scenario.  If the background uncertainties can be reduced to purely statistical, we estimate that a large portion of the interesting parameter space can be covered.   

Finally we considered  the possibility of observing this scenario by measuring the
$m_H$ and $m_A$ masses in the transverse mass, the collinear mass and an 
analytical solution based on the trial masses in the $2\tau_h$ channel.  We found that this
analytical solution allows for a greater number of events with physical solutions than the 
collinear method, while maintaining a similar resolution. The analytical method can also reconstruct larger values of the $H$ mass, which may be problematic using the transverse mass distribution, as the fall off of the transverse mass distribution is much softer for heavier masses.

{\em Note Added:  A new CMS multilepton analysis was recently presented at the HCP 2012 conference using 9.2 fb$^{-1}$ of 8 TeV data \cite{CMS-PAS-SUS-12-026}.  Due to an increase in the $p_T$ thresholds of taus, the observed number of events has decreased, which will have a strong impact on the efficiency of our signal. }

{\bf \noindent Acknowledgements}: We thank R.~Dermisek and S.~Samolwar  for useful discussions.  AM is supported by the U.S. Department of Energy under Contract No. DE-FG02-96ER40969.

\appendix

\section{NMSSM Realization of a large $H \to Z A$} \label{raxion}

In this section we provide an analytic explanation for the enhanced $H \to Z A$ branching 
ratio in the benchmark points in Tab.~\ref{tab:bmpara}, \ref{tab:bmspec}, \ref{tab:h1br}
and \ref{tab:h2br}. 
The minimization conditions for the super potential in Eq.~(\ref{eq:superpotential}) and the 
soft terms in Eq.~(\ref{eq:vsoft}) can be used to eliminate some of the soft SUSY breaking 
parameters:
\bea
m_{H_d}^2 &=& -\frac{\lambda}{2} (s^2 + v^2 s_\beta^2) + \frac{\lambda \kappa}{2} 
s^2 t_\beta - \frac{m_Z^2}{2} c_{2\beta} + m_\lambda s t_\beta \\
m_{H_u}^2 &=& - \frac{\lambda}{2} (s^2 + v^2 c_\beta^2) + \frac{\lambda \kappa}{2}
s^2 t_\beta^{-1} + \frac{m_Z^2}{2} c_{2\beta} + m_\lambda s t_\beta^{-1} \\
m_S^2 &=& - \frac{\lambda}{2} v^2 + \frac{\lambda \kappa}{2} 
v^2 s_{2\beta} - \kappa^2 s^2 + \frac{m_\lambda v^2}{2s} + m_\kappa  s
\eea
After substituting these solutions
into the scalar potential, there are six remaining free parameters $\lambda, 
\kappa, \beta, m_\lambda, m_\kappa$ and $s$. In terms of them, the tree-level CP-even mass matrix in the $(h_v^0,H_v^0, h_s^0)$ basis is
\bea
\mathcal{M}_{H^0}^2 = v^2 \left(\begin{array}{ccc}
r + \frac{M_Z^2}{v^2} & r\, {\rm cot} 2\beta & \lambda^2 \frac{s}{v} - R \\
r \, {\rm cot}2\beta & -r + \frac{\lambda \kappa s^2+2m_\lambda s}{v^2 \sin 2\beta}
& -R\, {\rm cot} 2\beta \\
\lambda^2 \frac{s}{v} - R & -R\, {\rm cot} 2\beta & \frac{2\kappa^2 s^2}{v^2} +
s \left(\frac{m_\lambda}{2s^2} - \frac{m_\kappa}{v^2} \right) 
\end{array}\right)
\eea
where 
\bea
r \equiv \left(\frac{\lambda^2}{2} - \frac{M_z^2}{v^2} \right)\sin^2 2\beta, \\
R \equiv \frac{1}{v} (2\lambda \kappa s + m_\lambda) \sin 2\beta.
\eea
Similarly, the CP-odd states $(A_v^0, A_s^0)$ have a tree-level mass matrix
\bea
\mathcal{M}_A^2 = \left(\begin{array}{cc}
(\lambda \kappa s^2 + 2 m_\lambda s) s_{2\beta}^{-1} & -v(\lambda \kappa s - 
m_\lambda) \\
-v(\lambda \kappa s - m_\lambda) & \left(\lambda \kappa + \frac{m_\lambda}{2s}
\right) v^2 s_{2\beta} + 3 s m_\kappa
\end{array} \right) \label{eq:mA2mat}
\eea
where the rotation angle $\theta_A$ satisfies
\bea
\tan 2\theta_A = \frac{4vs(\lambda \kappa s - m_\lambda)\sin 2\beta}{v^2
\sin^2 2\beta(2\lambda \kappa s + m\lambda) - 2s^2(\lambda \kappa s +2 m_\lambda
-3m_\kappa \sin 2\beta)}.
\eea
In this basis the $ZA$ couplings to the CP-even states have the form
\bea
\mathcal{L}_{\rm Higgs}^{\rm Kin} &=& D^\mu H_u^\dagger D_\mu H_u  + D^\mu H_d^\dagger D_\mu H_d \\
&\subset& -\frac{g_2}{2c_{\theta_W}} Z^\mu A_v^0 \overleftrightarrow{\partial_\mu}
 \left( s_{2\beta} h_v^0 + c_{2\beta} H_v^0 \right) \\
&=& -\frac{g_2}{2c_{\theta_W}} Z^\mu (c_{\theta_A} A_1^0 - s_{\theta_A} A_2^0) 
\overleftrightarrow{\partial_\mu}
 \left( s_{2\beta} h_v^0 + c_{2\beta} H_v^0 \right) \label{eq:lhza}
\eea
where $a \overleftrightarrow{\partial_\mu} b = a \partial_\mu b - b \partial_\mu 
a$ and in the last line we have rotated into the CP-odd Higgs mass basis. 

Within the NMSSM, a light pseudo-scalar can exist either in the Peccei-Quinn (PQ)
or the  $U(1)_R$ limit. 
If we take the $U(1)_R$ axion limit discussed in Ref.~\cite{Dobrescu:2000yn},
where $\mathcal{O}(10^{-3})\lsim m_{\lambda , \kappa}/v \ll 1 $,
the mass of the lightest CP-odd scalar is 
\bea
m_{A_1^0} \simeq \sqrt{3s} \left(m_\kappa \sin^2 \theta_A + 3 \frac{m_\lambda 
\cos^2 
\theta_A}{2\sin 2\beta} \right)^{1/2} + \mathcal{O}\left(\sqrt{\frac{m_{\lambda , 
\kappa}^3}{v}}\right)
\eea
and the rotation angle is
\bea
\tan \theta_A \simeq \frac{s}{v\sin 2\beta} + \mathcal{O}\left(\frac{m_{\lambda ,
\kappa}}{v}\right). \label{eq:cthetaAapprox}
\eea
In order to have a light $A^0_1$  with a large $g_{H_2^0ZA_1^0}$ coupling
we require large mixing in the CP-odd Higgs sector,  $\theta_A \simeq 
\frac{\pi}{4}$.   Also,  to satisfy charged Higgs limits from top decays \cite{Aad:2012tj, :2012cw}, we have to raise the $A_2$ mass since it is correlated with the charged Higgs.~\footnote{It is possible to avoid this constraint if the $H^\pm \to W^\pm A$ is kinematically allowed~\cite{Dermisek:2008uu}.  For that top quark cascade decay, the relevant limit is the following CDF study \cite{CDF10104}, which applies only for $m_A < 10$ GeV and has weaker limits.  We thank  R.~Dermisek for emphasizing this point to  us.}  In the R-axion limit, the magnitude of $m_{A_2^0}$ is set by
$\lambda \kappa (v^2 s_{2\beta} + s^2 s_{2\beta}^{-1})$, so the mass constraint implies that $\lambda$ and 
$\kappa$ are both order one.  To suppress the $g_{b\bar bH_2^0}$ coupling 
relative to $g_{H_2 ZA_1^0}$ we need to be at low $\tan \beta$. Additionally,
maximal mixing in the pseudo-scalar sector along with 
Eq.~(\ref{eq:cthetaAapprox}) suggests that $s \approx v \sin 2\beta \approx v$.
Large $g_{H_2^0 Z A_1^0}$ couplings also typically lead to an 
enhancement of the $g_{H_1^0 A_1^0 A_1^0}$ which reduces the $H_1$'s branching ratio into Standard Model decays, which is constrained by the
present Higgs boson measurements at the LHC~\cite{ATLASHiggs,CMSHiggs}. 
Hence, when $m_{A_1} < m_{H_1}/2$, viable benchmark points need a tuning so that the  $g_{H_1^0 A_1^0 A_1^0}$ coupling is suppressed, such that the branching ratios of $H_1$ remain SM-like.  

We note in passing that the PQ-axion is not a useful limit for our benchmarks.  In the PQ limit with $\kappa, 
m_\kappa$ small,  we can neglect the $\kappa$ and $m_\kappa$ terms in Eq.~(\ref{eq:mA2mat})
and find that $\theta_A \simeq \pi/4$ implies $s \simeq v s_{2\beta}/2$.
Since the VEV $s\sim v/2 \sim 125$ GeV is smaller than in the $PQ$ limit, the results small $\mu_{eff}$ often leads  to  chargino masses in violation of LEP2 bounds.  

\section{Neutrino Solution Using Trial Values of $m_H$ and $m_A$} \label{sec:mhmatest}

In this appendix, we show how to solve for the two neutrino four vectors $p_{\nu_{1,2}}$, knowing
the momenta of the visible decay products of the $\tau$'s
$p_{V_{1,2}}$ and the reconstructed $Z$ vector $p_Z$. There are eight
kinematic constraints
\bea
& & p_{\nu_1}^2 = 0 = p_{\nu_2}^2 \\
& & \left(p_{\nu_1} + p_{V_1} \right)^2 =  m_\tau^2 =  \left(p_{\nu_2} + 
p_{V_2} \right)^2 \\
& & m_A^2 = \left(p_{\nu_1} + p_{V_1} + p_{\nu_2} + 
p_{V_2} \right)^2 \\
& & m_H^2 = \left(p_Z + p_{\nu_1} + p_{V_1} + p_{\nu_2} + 
p_{V_2} \right)^2 \\
& & p_{\nu_1}^x + p_{\nu_2}^x = p_+^x \\
& & p_{\nu_1}^y + p_{\nu_2}^y = p_+^y
\eea
where  $m_{H,A}$ are
the trial masses and for simplicity we have defined
\bea
p_\pm = p_{\nu_1} \pm p_{\nu_2}.
\eea
In particular, the $x,y$ components of $p_+$ are the missing transverse energy components. 

The undetermined components of $p_+$ satisfy
\bea
& & p_+^z = \frac{E_Z E_+ - \Delta_H}{p_Z^z} \\
& & E_+^2 - \left(\frac{E_Z E_+ - \Delta_H}{p_Z^z}\right)^2 + 2 
\left(E_+ E_V - \frac{E_Z E_+ - \Delta_H}{p_Z^z} p_V^z \right) = \Delta_A
\label{eq:Eplus}
\eea
where
\bea
p_V &=& p_{V_1} + p_{V_2} \\
\Delta_H &=& \frac{1}{2}\left(m_H^2 - m_Z^2 - m_A^2\right) - 
p_Z \cdot p_V + p_Z^T \cdot p_+^T \\
\Delta_A &=& m_A^2 - p_V^2 + (p_+^T)^2 + 2 p_+^T \cdot p_V^T.
\eea
Therefore for a particular $(m_H,m_A)$, a physical solution is possible only 
if a real positive root to Eq.~(\ref{eq:Eplus}) exists.

The $p_-$ equations are
\bea
& & p_- \cdot p_{V_1} = \Delta_1 \\
& & p_- \cdot p_{V_2} = -\Delta_2 \\
& & p_- \cdot p_+ = 0 \\
& & p_-^2 = - p_+^2 \label{eq:Eminus}
\eea
where the terms on the right hand side are
\bea
\Delta_i = m_\tau^2 - p_{V_i}^2 - p_+ \cdot p_{V_i}
\eea
We can solve for the spatial components in terms of $E_-$
\bea
\vec{p}_- = E_- \vec{A} + \vec{B}
\eea
where
\bea
\vec{A} &=& X \left(\begin{array}{c}
E_{V_1}\\
E_{V_2}\\
E_+
\end{array} \right) \\
\vec{B} &=& X \left(\begin{array}{c}
-\Delta_1\\
\Delta_2\\
0
\end{array} \right) \\
X &=& \left(\begin{array}{ccc}
p_{V_1}^x &  p_{V_1}^y & p_{V_1}^z \\
p_{V_2}^x &  p_{V_2}^y & p_{V_2}^z \\
p_+^x & p_+^y &  p_+^z
\end{array} \right)^{-1}
\eea
Plugging this solution back into Eq.~(\ref{eq:Eminus}) we see
that a solution for the trial $m_H$ and $m_A$ values exists only if 
$E_-$ is real. Hence the conditions for a particular event arising from a 
trial $m_H, m_A$ are that at least one of the roots of Eq.~(\ref{eq:Eplus}) 
is positive and real and the $p_+$ corresponding to that root also satisfies
the condition
\bea
(\vec{A}(p_+) \cdot \vec{B}(p_+))^2 - (1 - |\vec{A}(p_+)|^2) 
( p_+^2 - |\vec{B}(p_+)|^2) &\geq& 0 \nonumber
\eea
which ensures that $E_-$ is real.

\bibliographystyle{JHEP}
\bibliography{multilepton}

\end{document}